\documentclass[english]{article}
\usepackage[T1]{fontenc}
\usepackage[latin9]{inputenc}
\usepackage{amsmath}
\usepackage{amssymb}
\usepackage[greek,british]{babel}
\begin{document}

\title{Quantum entanglement as an aspect of pure spinor geometry}

\author{V. Kiosses\medskip \\Theoretical Physics Department\\Aristotle
University of Thessaloniki\\54124 Thessaloniki, Greece\\bkios@physics.auth.gr}
\maketitle
\begin{abstract}
Relying on the mathematical analogy of the pure states of a two-qubit
system with four-component Dirac spinors, we provide an alternative
consideration of quantum entanglement using the mathematical formulation
of Cartan's pure spinors. A result of our analysis is that the Cartan
equation of two qubits state is entanglement sensitive in a way that
the Dirac equation for fermions is mass sensitive. The Cartan equation
for unentangled qubits is reduced to a pair of Cartan equations for
single qubits as the Dirac equation for massless fermions separates
into two Weyl equations. Finally, we establish a correspondance between
the separability condition in qubit geometry and the separability
condition in spinor geometry.
\end{abstract}

\section{Introduction}

Although in 1932 von Neumann had completed basic elements of nonrelativistic
quantum description of the world, it were Einstein, Podolsky and Rosen
(EPR) and Sch$\ddot{o}$dinger who first recognized a ``spooky''
feature of quantum machinery which lies at center of interest of physics
of XXI century. This feature, known as entanglement \cite{Schrodinger},
implies the existence of global states of composite system which cannot
be written as a product of the states of individual subsystems \cite{Poland}.
The definition of entanglement relies on the tensor-product structure
of the state-space of a composite quantum system \cite{Peres}. In system of indistinguishable
particles, quantum statistics applies
and therefore their state-space is not naturally endowed
with a tensor product structure \cite{Zanardi}. Recently an algebraic approach to quantum
non - separability was applied to the case of two qubits, based on
the partition of the algebra of observables into independent subalgebras
\cite{Derkacz}. In this paper we shall tackle quantum entanglement
from a geometric perspective.

The concept of quantum entanglement is believed to play an essential
role in quantum information processing. Due to the fundamental significance
of quantum information theory, many efforts have been devoted to the
characterization of entanglement in terms of several physically equivalent
measures, such as entropy and concurrence \cite{Wootters,Vedral et al.}.
One of the most fundamental and most studied models of quantum information
theory is the two-qubit (two-spin-$\tfrac{1}{2}$) system (see, e.g.,
\cite{non-local operations}). The pure states of the two-qubit system
are described in terms of vectors of a four-dimensional Hermitian
(complex) vector space $\mathbb{C}_{4}$. It is of interest to describe
their quantum evolution in a suitable representation space, in order
to get some insight into the subtleties of this complicated problem.
A well known tool in quantum optics for the single qubit is the Bloch
sphere representation, where the simple qubit state is faithfully
represented, up to its overall phase, by a point on a standard sphere
$S^{2}$, whose coordinates are expectation values of physically interesting
operators for the given quantum state. Guided by the relation between
the Bloch sphere and a geometric object called the Hopf fibration
of the $S^{3}$ hypersphere \cite{Urbanke}, a generalization for
a two-qubit system was proposed \cite{Mosseri - Dandoloff}, in the
framework of the (high dimensional) $S^{7}$ sphere Hopf fibration,
and will be recalled below.

In the last few years the application of quantum information principles
and viewpoints within relativistic quantum mechanics and quantum field
theory has begun to be explored \cite{Peres et al.,Preskill}. In
the realm of relativistic quantum mechanics and quantum field theory
in 3+1-dimensional Minkowski space-time fermions, an essential part
of physical reality, are mainly described by means of Dirac spinors
\textendash{} elements of a four-component complex vector space which
are transforming under the spin representation of the Poincar\'{e}
group (inhomogeneous Lorentz group) of space-time. Relying on this
mathematical analogy of the pure states of the two-qubit system of
quantum information theory with four-component spinors we introduce
the concept of spinors entanglement and explore the connection between
one of the most fundamental models of relativistic quantum mechanics
and one of the most fundamental models of quantum information theory.
In this paper certain aspects of this problem are studied. The same
problem, but from a different viewpoint, is addressed by two different papers 
\cite{Rajagopal,Scharnhorst}.

As far as computation is concerned, the pure spinors provide us
(by nature) a natural geometrical interpretation. Spinors were first
used under such a name by physicists in the field of quantum mechanics.
In most of their general mathematical forms, spinors were discovered
in 1913 by Elie Cartan \cite{Cartan}, who conjectured that the fundamental
geometry appropriate for the description and understanding of elementary
natural phenomena is spinor geometry, more precisely the geometry
of simple spinors, later named pure by C. Chevalley \cite{Chevalley},
rather than the one of euclidean vectors which may be constructed
bilinearly from spinors. A fundamental property of pure spinors is
that of generating null vectors in momentum spaces, where the Cartan
equations defining pure spinors are identical to equations of motion
for massless systems \cite{Budinich}. In this paper, we find that
a two-component pure spinor, naturally generating the expectation values of Pauli operators, may be associated
with a simple qubit state. Adopting the constructive formula of imbedding spinor spaces in higher dimensional ones
from pure spinor theory \cite{Budinich}, we first build up a four-component spinor and then a eight-component ones. 
 A result of our analysis is that the Cartan's
equation with four-component spinor, which is associated to two qubits state, is entanglement sensitive in a similar
way to the Dirac equation, for massless fermions, separates into two
equations (the Cartan-Weyl equations).

The paper is organised as follows. In section 2, after a short introduction to the properties of pure spinors
we consider the problem of imbedding spinor spaces and null vector spaces in higher dimensional ones, 
and we show how it may be easily solved, in the case of pure spinors, with the use of two propositions. In section 3,
using the Bloch sphere representation, we derive the Cartan's equations of a single qubit
state. In section 4, we apply the aforementioned technique of imbedding spinor spaces in higher dimensional ones 
in order to obtain Cartan's equations
of two-qubits state. This leads to a discussion of the
role of entanglement in structure of pure spinor geometry. In section
5 we consider the seperability conditions of two qubits in pure spinor
geometry. We establish, in section 6, a correspondance between the
separability condition in qubit geometry and the separability condition
in spinor geometry. In section 7 we generalize the results from two qubits to four qubits. Finally, we conclude
in section 8 with a brief discussion.

\section{An introduction to pure spinor geometry}\label{introd. pure spinor}

Spinors were first used under such a name by physicists in the field of quantum
mechanics. In most of their general mathematical forms, spinors were discovered
in 1913 by Elie Cartan \cite{Cartan}. An appropriate instrument for dealing with spinors
is represented by Clifford algebras. We will present here a short review of the
needed formalism. For more on the subject we refer to \cite{Coquereax,Budinich-Trautman,Trautman}.

Given a $2n$ dimensional real vector space $V=\mathbb{R}^{2n}$ with a quadratic form $h$ of signature $(k,l)$, $k+l=2n$
and the corresponding Clifford algebra 
$\mathcal{C}l\left(2n\right)$ generated by a unit $1$ and $2n$ elements $\gamma_1 \ldots \gamma_{2n}$ satisfying:
\begin{equation}
 \left\{\gamma_\mu,\gamma_\nu\right\}=2h_{\mu\nu},
\end{equation}
a spinor $\psi$ is a vector of the $2^n$ dimensional representation space $S$ of 
$\mathcal{C}l\left(2n\right)=End\,S$, defined by:
\begin{equation}
 x\,\psi=x^\mu\gamma_\mu\psi=0,\qquad \mu=1,\ldots,2n\label{eq:1.2}
\end{equation}
where $x^\mu$ are the orthogonal components of a vector $x\in V$. Eq.(\ref{eq:1.2}) is known as the Cartan's equation.
Given $\psi\neq0$, the fundamental quadratic form of $V$ is $x^\mu x_\mu=0$. Therefore we may define $T_d(\psi)$,
\begin{equation}
 T_d(\psi)=\left\{x\in V|x^\mu\gamma_\mu\psi=0\right\},
\end{equation}
as the $d$-dimensional subspace of $V$ associated with $\psi$. It is easily seen that $T_d(\psi)$ is totally null. A spinor $\psi$
is said to be pure if $T_d(\psi)$ is a maximal, i.e. $d=n$, and totally null subspace of $V$.

Let $\gamma_{2n+1}:=\gamma_1\gamma_2\ldots\gamma_{2n}$ represents the volume element of $\mathcal{C}l\left(2n\right)$,
then the spinors $\varphi_W^{\pm}$ obeying
\begin{equation}
 \gamma_{2n+1}\varphi_W^{\pm}=\pm\varphi_W^{\pm}
\end{equation}
are named Weyl spinors and for them the defining equation (\ref{eq:1.2}) becomes
\begin{equation}
 x^{\mu}\gamma_\mu\left(1\pm\gamma_{2n+1}\right)\psi=0,
\end{equation}
because it holds
\begin{equation}
 \varphi_W^{\pm}=\frac{1}{2}\left(1\pm\gamma_{2n+1}\right)\psi.
\end{equation}
For $n\leq3$ all Weyl spinors $\varphi_W^{\pm}$ are pure \cite{Budinich}. 

The $2^{n-1}$ dimensional spaces $S_{+}$ and $S_{-}$ of Weyl spinors are endomorphism spaces of the even Clifford 
subalgebra $\mathcal{C}l_0\left(2n\right)$ of $\mathcal{C}l\left(2n\right)$ that is:
\begin{equation}
 \mathcal{C}l_0\left(2n\right)=End\,S_{\pm}.\label{eq:1.7}
\end{equation}
We have further:
\begin{equation}
\psi=\varphi_{W}^{+}\oplus\varphi_{W}^{-},\quad S=S_{+}\oplus S_{-}\label{eq:1.8}
\end{equation}
and $\mathcal{C}l\left(2n\right)=2\,\mathcal{C}l_0\left(2n\right)$.

These definitions may easily be extended also to odd dimensional spaces,
in fact, since the volume element $\gamma_{2n+1}$, for every $\gamma_{a}$,
obey to:$\left\{ \gamma_{a},\gamma_{2n+1}\right\} =2\delta_{a,2n+1}$, $1\leq a \leq 2n$,
we have that $\gamma_{1},\gamma_{2},\ldots,\gamma_{2n},\gamma_{2n+1}$
generate the Clifford algebra $\mathcal{C}l\left(2n+1\right)$ and there is the isomorphism \cite{Budinich-Trautman}:
\begin{equation}
 \mathcal{C}l_0\left(2n+1\right)\simeq \mathcal{C}l\left(2n\right)\label{eq:1.9}
\end{equation}
both being central simple. 
The corresponding $2^{n}$ component spinors are called Pauli spinors
for $\mathcal{C}l_{0}\left(2n+1\right)$ and
Dirac spinors for $\mathcal{C}l\left(2n\right)$. $\mathcal{C}l\left(2n+1\right)$ instead is not central simple and there is
the isomorphism \cite{Budinich-Trautman}:
\begin{equation}
 \mathcal{C}l\left(2n+1\right)\simeq \mathcal{C}l_0\left(2n+2\right).\label{eq:1.10}
\end{equation}
For embedding spinors in higher dimensional spinors we may then use eqs.(\ref{eq:1.7}),(\ref{eq:1.8}),(\ref{eq:1.9}) and
(\ref{eq:1.10}):
\begin{equation}
 \mathcal{C}l\left(2n\right)\simeq \mathcal{C}l_0\left(2n+1\right)\hookrightarrow \mathcal{C}l\left(2n+1\right)
 \simeq \mathcal{C}l_0\left(2n+2\right)\hookrightarrow \mathcal{C}l\left(2n+2\right)
\end{equation}
corresponding to the spinor embeddings:
\begin{equation}
 \psi_D\simeq\psi_P\hookrightarrow\psi_P\oplus\psi_P\simeq\varphi_W^{+}\oplus\varphi_W^{-}=\Psi=\psi_D\oplus\psi_D
\end{equation}
which means that a $2^n$ components Dirac spinors may be considered equivalent to a doublet of $2^{n-1}$ components
Weyl, Pauli or Dirac spinors.\\
\underline{\emph{Proposition 1}} Given a real space $V=\mathbb{R}^{2n}$ with its Clifford algebra $\mathcal{C}l\left(2n\right)$, with generators 
$\gamma_a$, let $\psi$ represents one spinor of the endomorphism spinor space of $\mathcal{C}l\left(2n\right)$ and 
$\bar{\psi}$ its adjoint, which is given as $\bar{\psi}=\psi^{\dagger}\gamma_h\gamma_k\ldots$ for $h,k,\ldots$ the timelike dimensions.
Then, the vector $x\in V$, with components:
\begin{equation}
x_a=\bar{\psi}\gamma_a \psi\label{eq:1.13}
\end{equation}
is null if and only if $\psi$ is a pure spinor. The proof is given in reference \cite{Budinich2}.

Let us now proceed with the imbedding of spinor spaces and null-vector spaces in higher dimensional ones. In order to
do this we adopt Proposition 1 and exploit the properties of pure spinors. Let us consider $\mathcal{C}l\left(2n\right)=End\,S$
and $\psi\in S$, then for $\varphi_{\pm}=\tfrac{1}{2}\left(1\pm\gamma_{2n+1}\right)\psi$ pure, the vectors $x^{\pm}$ with
components
\begin{equation}
 x_a^{\pm}=\bar{\psi}\gamma_{a}\left(1\pm\gamma_{2n+1}\right)\psi,\qquad a=1,2,\ldots,2n\label{eq:1.14}
\end{equation}
are of the form (\ref{eq:1.13}) and therefore are null:
\begin{equation}
 x_a^{\pm}x^a_{\pm}=0.
\end{equation}
Let us sum them and we obtain:
\begin{equation}
 x_a^{+}+x_a^{-}=X_a=\bar{\psi}\gamma_{a}\psi,\qquad a=1,2,\ldots,2n,\label{eq:1.16}
\end{equation}
where $\psi=\varphi_{+}\oplus\varphi_{-}\in S$, while $\varphi_{\pm}\in S_{\pm}$, where $\mathcal{C}l_0\left(2n\right)=End\,S_{\pm}$ 
and therefore
\begin{equation}
 S=S_{+}\oplus S_{-}.
\end{equation}
The corresponding operation in vector space is given by proposition:\\
\underline{\emph{Proposition 2}} Given two pure spinors $\varphi_{\pm}$ of $\mathcal{C}l_0\left(2n\right)$ and the corresponding 
null vectors $x_{\pm}\in \mathbb{R}^{2n}$ with components
\begin{equation}
 x_a^{\pm}=\bar{\varphi}_{\pm}\gamma_{a}\varphi_{\pm}
\end{equation}
their sum
\begin{equation}
 x_a^{+}+x_a^{-}=X_a=\bar{\psi}\gamma_{a}\psi,
\end{equation}
where $\psi=\varphi_{+}\oplus\varphi_{-}$, is a projection on $\mathbb{R}^{2n}$ of a null vector $X\in \mathbb{R}^{2n+2}$
which is obtained by adding to $X_a$ the two extra components
\begin{equation}
 X_{2n+1}=\bar{\psi}\gamma_{2n+1}\psi\qquad X_{2n+2}=\pm i\bar{\psi}\psi
\end{equation}
provided $\psi$ satisfies the conditions of purity as a Weyl spinor of $\mathcal{C}l_0\left(2n+2\right)$.

In this way we see that to the direct sum of pure spinor spaces, giving rise to a higher dimensional pure spinor 
space $S=S_{+}\oplus S_{-}$, where $\mathcal{C}l_0\left(2n\right)=End\,S_{\pm}$ and $\mathcal{C}l\left(2n+2\right)=End\,S$,
there correspond the sum of null vectors giving rise to a higher dimensional null vector:
\begin{equation}
 x^{+}+x^{-}\hookrightarrow X
\end{equation}
where $x^{\pm}\in \mathbb{R}^{2n}$ and $X\in \mathbb{R}^{2n+2}$; and we have an instrument for imbedding spinor spaces 
in higher dimensional spinor spaces and correspondingly null vector spaces in higher dimensional null vector spaces. 
This instrument we are using to build a four-component spinor and eight-component spinor, which, as we will see, are 
associated to two-qubits and four-qubits entangled states respectively.

\section{The Cartan equation of single qubit state}

A single qubit state reads
\begin{equation}
\left|\psi\right\rangle =a\left|0\right\rangle +b\left|1\right\rangle ,\qquad a,b\in\mathbb{C}.
\end{equation}
In the spin $\tfrac{1}{2}$ context, the orthonormal basis $\left\{ \left|0\right\rangle ,\left|1\right\rangle \right\} $
is composed of the two eigenvectors of the $\sigma_{z}$ Pauli spin
operator. A conveniant way to represent $\left|\psi\right\rangle $
(up to a global phase) is provided by the Bloch sphere. The set of
states $\exp i\varphi\left|\psi\right\rangle $, with $\varphi\in\left[0,2\pi\right[$,
is mapped onto a point on $S^{2}$ (the usual sphere in $\mathbb{R}^{3}$):
\begin{equation}
X_{1}^{2}+X_{2}^{2}+X_{3}^{2}=X_{0}^{2}\label{eq:2.2}
\end{equation}
with coordinates
\begin{eqnarray}
X_{0}=\left\langle \psi\right. \left|\psi\right\rangle  & = & \left|a\right|^{2}+\left|b\right|^{2},\nonumber \\
X_{1}=\left\langle \psi\right|\sigma_{1}\left|\psi\right\rangle  & = & 2\text{Re}\left(a^{*}b\right),\nonumber \\
X_{2}=\left\langle \psi\right|\sigma_{2}\left|\psi\right\rangle  & = & 2\text{Im}\left(a^{*}b\right),\label{eq:9}\\
X_{3}=\left\langle \psi\right|\sigma_{3}\left|\psi\right\rangle  & = & \left|a\right|^{2}-\left|b\right|^{2},\nonumber 
\end{eqnarray}
($a^{*}$ is the complex conjugate of $a$) to be the expectation
values of physically interesting operators for the given quantum state.
Here, $\sigma_{1,2,3}$ are the standard Pauli matrices:
\begin{equation}
\sigma_{1}=\left(\begin{array}{cc}
0 & 1\\
1 & 0
\end{array}\right),\quad\sigma_{2}=\left(\begin{array}{cc}
0 & -i\\
i & 0
\end{array}\right),\quad\sigma_{3}=\left(\begin{array}{cc}
1 & 0\\
0 & -1
\end{array}\right).
\end{equation}
Recall the relation between Bloch sphere coordinates and the pure
state density matrix $\rho_{\left|\psi\right\rangle }$:
\begin{equation}
\rho_{\left|\psi\right\rangle }=\rho_{\exp i\varphi\left|\psi\right\rangle }=\left|\psi\right\rangle \left\langle \psi\right|=\frac{1}{2}\left(\begin{array}{cc}
1+X_{3} & X_{1}-iX_{2}\\
X_{1}+iX_{2} & 1-X_{3}
\end{array}\right).
\end{equation}
The vanishing determinant of $\rho$ is related to the radius, $X_0$,
of the Bloch sphere.
For a normalized state $\left|\psi\right\rangle$ with
\begin{equation}
\left|a\right|^{2}+\left|b\right|^{2}=1,\label{eq:2.5}
\end{equation}
eq.(\ref{eq:2.2}) becomes
\begin{equation}
X_{1}^{2}+X_{2}^{2}+X_{3}^{2}=1.\label{2.6}
\end{equation}

We now proceed to the linearization of eq.(\ref{eq:2.2}). Our procedure
is exactly the same as followed by Dirac in his classical paper on
the spinning electron\cite{Dirac} and is based on real Clifford algebras
\cite{Traubenberg}. The coordinates of the Bloch sphere in eq.(\ref{eq:2.2}),
are associated with Clifford algebra $\mathcal{C}l\left(3\right)$, generated by
the Pauli matrices $\sigma_{1},\sigma_{2},\sigma_{3}$, and can be
written as\footnote{Einstein notation is adopted to indicate summation, $X_{i}\sigma^{i}=\sum_i X_{i}\sigma_{i}$}
\begin{equation}
X=X_{i}\sigma^{i}.
\end{equation}
Following Cartan we may define a spinor $\left|\psi\right\rangle $
with the equation (terms with repeated indices are meant to be summed):
\begin{equation}
\left(X_{i}\sigma^{i}-X_0\right)\left|\psi\right\rangle =0,\quad i=1,2,3.\label{eq:14}
\end{equation}
This equation is the linearised version of the quadratic equation
(\ref{eq:2.2}). Notice that multiplying (\ref{eq:14}) from the left
with $\left\langle \psi\right|=\left|\psi\right\rangle ^{\dagger}$
and using eqs.(\ref{eq:9}) we obtain the Bloch sphere (\ref{eq:2.2}).

\section{The Cartan equation of two-qubits state}\label{two-qubits}

Now consider two ``separable'' single qubits, let say Alice:
\begin{equation}
\left|\psi^{A}\right\rangle =a\left|0\right\rangle +b\left|1\right\rangle ,\qquad a,b\in\mathbb{C}
\end{equation}
represented in Bloch sphere:
\begin{equation}
\left(X_{1}^{A}\right)^{2}+\left(X_{2}^{A}\right)^{2}+\left(X_{3}^{A}\right)^{2}=\left(X_{0}^{A}\right)^{2}
\end{equation}
and Bob:
\begin{equation}
\left|\psi^{B}\right\rangle =c\left|0\right\rangle +d\left|1\right\rangle ,\qquad c,d\in\mathbb{C}
\end{equation}
represented in Bloch sphere:
\begin{equation}
\left(X_{1}^{B}\right)^{2}+\left(X_{2}^{B}\right)^{2}+\left(X_{3}^{B}\right)^{2}=\left(X_{0}^{B}\right)^{2}.
\end{equation}
If we exploit the fact that the Bloch sphere coordinates, $X_i^{A,B}$, are defined
uniquely up to sign, the associated Cartan's equations take the form:
\begin{equation}
\begin{aligned}\left(X_{i}^{A}\sigma^{i}-X_{0}^{A}\right)\left|\psi^{A}\right\rangle  & =0\\
\left(X_{i}^{B}\sigma^{i}+X_{0}^{B}\right)\left|\psi^{B}\right\rangle  & =0
\end{aligned}
\label{eq:19}
\end{equation}
respectively. With ``separable'' we mean that the associated Cartan's equations of the corresponding qubits
are decoupled. 

Equations (\ref{eq:19}) may be expressed as a single equation for
the four component Dirac spinor $\left|\Psi\right\rangle =\left|\psi^{A}\right\rangle \oplus\left|\psi^{B}\right\rangle $.
Indicating the matrices $\gamma^{\mu}$ as
\begin{equation}
\gamma^{\mu}:\,\gamma^{i}=\left(\begin{array}{cc}
0 & \sigma^{i}\\
\sigma^{i} & 0
\end{array}\right),\;\gamma^{0}=\left(\begin{array}{cc}
0 & \mathbf{1}_{2}\\
-\mathbf{1}_{2} & 0
\end{array}\right),\label{eq:20}
\end{equation}
with
\begin{equation}
\left\{ \gamma^{\mu},\gamma^{\nu}\right\} =2\eta^{\mu\nu}=2\left(\begin{array}{cccc}
-1 & 0 & 0 & 0\\
0 & 1 & 0 & 0\\
0 & 0 & 1 & 0\\
0 & 0 & 0 & 1
\end{array}\right)\label{eq:21}
\end{equation}
and 
\begin{equation}
\gamma^{5}=-i\gamma^{0}\gamma^{1}\gamma^{2}\gamma^{3}=\left(\begin{array}{cc}
\mathbf{1}_{2} & 0\\
0 & -\mathbf{1}_{2}
\end{array}\right)
\end{equation}
 its volume element, we may write (\ref{eq:19}) in the form
\begin{equation}
X_{\mu}^{A,B}\gamma^{\mu}\left(1\pm\gamma^{5}\right)\left|\Psi\right\rangle =0\label{eq:23}
\end{equation}
 where $\left|\psi^{A,B}\right\rangle =\frac{1}{2}\left(1\pm\gamma^{5}\right)\left|\Psi\right\rangle $
the Weyl spinors and eigenspinors of $\gamma^{5}$. Writing the Hermitean
conjugate of eq.(\ref{eq:23}):
\begin{equation}
X_{\mu}^{A,B}\left\langle \bar{\Psi}\right|\gamma_{\mu}\left(1\pm\gamma^{5}\right)=0
\end{equation}
where $\left\langle \bar{\Psi}\right|=\left\langle \Psi\right|\gamma^{0}$,
we may define the quantities $X_{\mu}^{A,B}$ by the relations%
\footnote{remember that $\left(\frac{1\pm\gamma^{5}}{2}\right)^{2}=\left(\frac{1\pm\gamma^{5}}{2}\right)$%
}
\begin{equation}
X_{\mu}^{A,B}=\left\langle \bar{\Psi}\right|\gamma_{\mu}\left(\frac{1\pm\gamma^{5}}{2}\right)\left|\Psi\right\rangle .
\end{equation}
It is not hard to check that:
\begin{equation}
X_{\mu}^{A,B}X_{A,B}^{\mu}=0\Rightarrow\left(X_{1}^{A,B}\right)^{2}+\left(X_{2}^{A,B}\right)^{2}+\left(X_{3}^{A,B}\right)^{2}=\left(X_{0}^{A,B}\right)^{2}
\end{equation}
which means that writing the equations (\ref{eq:19}) in a single equation doesn't affect the independence
of the corresponding Bloch spheres.

\subsection{Left-handed and right-handed Bloch spheres\label{sub:Left-handed-and-right-handed}}

Writing the two qubits in this way, an interesting aspect of the corresponding
Bloch spheres is revealed. An arbitrary unitary operation on the single
qubit can be represented by a unitary matrix on $\mathbb{C}^{2}$,
a member of $SU\left(2\right)$ group. The same operation can be also
written with a rotation matrix, from $SO\left(3\right)$ group, of
the Bloch sphere by an angle $\omega_{ij}$ about axis defined by
bivector $\sigma^{i}\wedge\sigma^{j}$. Thus under an orthogonal transformation
$L_{i}^{j}\in SO\left(3\right)$ of Bloch sphere coordinates a single
qubit state transforms according to
\begin{equation}
\left|\psi\right\rangle '\left(L_{i}^{j}X^{i}\right)=\Lambda\left|\psi\right\rangle \left(X^{j}\right)
\end{equation}
where 
\begin{equation}
\Lambda=\exp\left(\frac{1}{2}\omega_{ij}W^{ij}\right)=\exp\left(\frac{i}{8}\omega_{ij}\left[\sigma^{i},\sigma^{j}\right]\right)
\end{equation}
the invertible elements that define $Spin\left(3\right)$ group, the
double cover of $SO\left(3\right)$.

Writing the two single non-entangled qubits in terms of Weyl spinors,
eq.(\ref{eq:23}), as we have seen, the associated Clifford algebra
is generated by $\gamma^{\mu}$ satisfying (\ref{eq:21}). This representation
is reducible. Specially, in block notations let
\begin{equation}
\gamma^{\mu}=\left(\begin{array}{cc}
0 & \zeta^{\mu}\\
\theta^{\mu} & 0
\end{array}\right).
\end{equation}
Then $2\times2$ matrices $\eta^{\mu}$ and $\theta^{\mu}$ satisfy
\begin{equation}
\zeta^{\mu}\theta^{\nu}+\zeta^{\nu}\theta^{\mu}=2\eta^{\mu\nu},\qquad\theta^{\mu}\zeta^{\nu}+\theta^{\nu}\zeta^{\mu}=2\eta^{\mu\nu}.
\end{equation}
The matrices $M_{L}^{\mu\nu}\equiv\frac{i}{4}\left[\zeta^{\mu},\theta^{\nu}\right]$
and $M_{R}^{\mu\nu}\equiv\frac{i}{4}\left[\theta^{\mu},\zeta^{\nu}\right]$
generate different rotations in the $\mu-\nu$ plane. The group thus
has two ``pieces'' 
\begin{equation}
\begin{aligned}\Lambda_{L}= & \exp\left(\frac{1}{2}\omega_{\mu\nu}M_{L}^{\mu\nu}\right),\\
\Lambda_{R}= & \exp\left(\frac{1}{2}\omega_{\mu\nu}M_{R}^{\mu\nu}\right).
\end{aligned}
\end{equation}
They are topologically disjunct (disjoint), and there is no continuous
path from one piece to the other. So we can define two types of spinors
besides two kinds of generators $\eta^{\mu}$ and $\theta^{\mu}$.
Under an arbitrary unitary operation the pair of qubits $\left|\psi^{A}\right\rangle $
and $\left|\psi^{B}\right\rangle $ transforms as
\begin{eqnarray}
\left|\Psi\right\rangle '=\exp\left(\frac{1}{2}\omega_{\mu\nu}M^{\mu\nu}\right) & = & \exp\left(\frac{i}{8}\omega_{\mu\nu}\left[\gamma^{\mu},\gamma^{\nu}\right]\right)\left|\Psi\right\rangle \nonumber \\
\left(\begin{array}{c}
\left|\psi^{A}\right\rangle '\\
\left|\psi^{B}\right\rangle '
\end{array}\right) & = & \left(\begin{array}{c}
\Lambda_{L}\left|\psi^{A}\right\rangle \\
\Lambda_{R}\left|\psi^{B}\right\rangle 
\end{array}\right)
\end{eqnarray}
identifying Alice qubit as left-handed Weyl spinor and Bob qubit as
right-handed Weyl spinor. So Cartan-Weyl equations (\ref{eq:23})
define two types of qubit Bloch spheres, a left-handed and a right-handed
Bloch sphere in correspondance to the left-handed and right-handed
Weyl spinor.

\subsection{Mixing terms and ``entanglement''}\label{mixing terms}

In order to couple these two Cartan-Weyl equations we take the sum of the components
of the two Bloch spheres 
\begin{equation}
X_{\mu}^{A}+X_{\mu}^{B}=X_{\mu}=\left\langle \bar{\Psi}\right|\gamma_{\mu}\left|\Psi\right\rangle ,\label{eq:33}
\end{equation}
where $\left|\Psi\right\rangle =\left|\psi^{A}\right\rangle \oplus\left|\psi^{B}\right\rangle $.
This means that starting from two pure spinor spaces we operated the
most obvious operation: their direct sum spanned by spinors of double
dimension. The corresponding operation in vector space, represented
by eq.(\ref{eq:33}), defines vector $X_{\mu}$, with components:
\begin{equation}
\begin{aligned}X_{1}=\left\langle \bar{\Psi}\right|\gamma_{1}\left|\Psi\right\rangle = & a^{*}b+b^{*}a-c^{*}d-d^{*}c\\
X_{2}=\left\langle \bar{\Psi}\right|\gamma_{2}\left|\Psi\right\rangle = & i\left(-a^{*}b+b^{*}a+c^{*}d-d^{*}c\right)\\
X_{3}=\left\langle \bar{\Psi}\right|\gamma_{3}\left|\Psi\right\rangle = & \left|a\right|^{2}-\left|b\right|^{2}-\left|c\right|^{2}+\left|d\right|^{2}\\
X_{0}=\left\langle \bar{\Psi}\right|\gamma_{0}\left|\Psi\right\rangle = & -\left|a\right|^{2}-\left|b\right|^{2}-\left|c\right|^{2}-\left|d\right|^{2}
\end{aligned}
\end{equation}
which is non null:
\begin{equation}
X_{1}^{2}+X_{2}^{2}+X_{3}^{2}-X_{0}^{2}=-C^{2}.\label{eq:35}
\end{equation}

According to proposition 2, given two pure spinors $\left|\psi^{A,B}\right\rangle $
and the corresponding null vectors $X_{\mu}^{A,B}\in\mathbb{R}^{4}$
with components
\begin{equation}
X_{\mu}^{A,B}=\left\langle \bar{\Psi}\right|\gamma_{\mu}\left(\frac{1\pm\gamma^{5}}{2}\right)\left|\Psi\right\rangle 
\end{equation}
their sum eq.(\ref{eq:33}) is a projection on $\mathbb{R}^{4}$ of
a null vector $X_{l}\in\mathbb{R}^{6}$ which is obtained by adding
to $X_{\mu}$ the two extra components
\begin{equation}
\begin{aligned}X_{4}= i\left\langle \bar{\Psi}\right|\left.\Psi\right\rangle =  & i\left(a^{*}c+b^{*}d-c^{*}a-d^{*}b\right),\\
X_{5}=\left\langle \bar{\Psi}\right|\gamma_{5}\left|\Psi\right\rangle = & -a^{*}c-b^{*}d-c^{*}a-d^{*}b.
\end{aligned}
\end{equation}
It is easy to verify that
\begin{equation}
X_{4}^{2}+X_{5}^{2}=C^{2},
\end{equation}
thus eq.(\ref{eq:35}) takes the form:
\begin{equation}
X_{1}^{2}+X_{2}^{2}+X_{3}^{2}+X_{4}^{2}+X_{5}^{2}-X_{0}^{2}=0,\label{eq:39}
\end{equation}
which may be associated with the generalized Bloch sphere, first introduced by Mosseri and Dandoloff \cite{Mosseri - Dandoloff}.
In that paper the authors focus on the projective Hilbert space, taking into account the global phase freedom. 
The single qubit Hilbert space is the unit sphere $S^3$. The corresponding projective Hilbert space is such that all 
states differing by a global phase are identified and corresponds to the well-known Bloch sphere, $S^2$. In order to 
generalize the Bloch sphere to the two-qubits projective Hilbert space, they used quaternions and found that a unit 
sphere, $S^{4}$, is the projective Hilbert space of the two-qubits Hilbert space, $S^7$ (for more about this subject 
I refer to \cite{Mosseri - Dandoloff}). As for the Bloch sphere case, the coordinates that define $S^4$ are also 
expectation values of operators in the two-qubits state. Setting these operators to be given by the matrices $\{\gamma_\mu,i\mathbf{1}_{4},\gamma_5\}$,
Eq.(\ref{eq:39}) may be considered as a representation of a generalized Bloch sphere.

With $X_{l}$ we may generate the Cartan's equation
\begin{equation}
\left(X_{\mu}\gamma^{\mu}+iX_{4}+X_{5}\gamma^{5}\right)\left|\Psi\right\rangle =0,
\end{equation}
or in block notation 
\begin{eqnarray}
\left(\begin{array}{cc}
\left(iX_{4}+X_{5}\right)\mathbf{1}_{2} & \left(X_{i}\sigma^{i}+X_{0}\right)\mathbf{1}_{2}\\
\left(X_{i}\sigma^{i}-X_{0}\right)\mathbf{1}_{2} & \left(iX_{4}-X_{5}\right)\mathbf{1}_{2}
\end{array}\right)\left(\begin{array}{c}
\left|\psi^{A}\right\rangle \\
\left|\psi^{B}\right\rangle 
\end{array}\right) & = & 0\Rightarrow\nonumber \\
\left\{ \begin{array}{c}
\left(X_{i}\sigma^{i}+X_{0}\right)\left|\psi^{B}\right\rangle =-\left(X_{5}+iX_{4}\right)\left|\psi^{A}\right\rangle \\
\left(X_{i}\sigma^{i}-X_{0}\right)\left|\psi^{A}\right\rangle =\left(X_{5}-iX_{4}\right)\left|\psi^{B}\right\rangle 
\end{array}\right. & \begin{array}{c}
\\
\\
\end{array} & \begin{array}{c}
\\
\\
\end{array}\label{eq:41}
\end{eqnarray}
Let us express the complex vector $X_{5}+iX_{4}$ in the polar form:
\begin{equation}
X_{5}\pm iX_{4}=M\, exp\left(\pm i\frac{\omega}{2}\right)
\end{equation}
Substituting this in eq.(\ref{eq:41}) and multiplying the second
by $exp\left(i\frac{\omega}{2}\right)$ we obtain: 
\begin{eqnarray}
\left\{ \begin{array}{c}
\left(X_{i}\sigma^{i}+X_{0}\right)\left|\psi^{B}\right\rangle =-M\, e^{i\frac{\omega}{2}}\left|\psi^{A}\right\rangle \\
\left(X_{i}\sigma^{i}-X_{0}\right)e^{i\frac{\omega}{2}}\left|\psi^{A}\right\rangle =M\left|\psi^{B}\right\rangle 
\end{array}\right.\label{eq:43}
\end{eqnarray}
We see that $\left|\psi^{A}\right\rangle $ appears with a phase factor
$e^{i\frac{\omega}{2}}$ corresponding to a rotation through an angle
$\omega$ in the circle
\begin{equation}
X_{4}^{2}+X_{5}^{2}=M^{2}.
\end{equation}
Note that setting 
\begin{equation}
X_{4}=X_{5}=0\Rightarrow M=0,\label{eq:45}
\end{equation}
eqs.(\ref{eq:41}) reduce to
\begin{eqnarray}
\left\{ \begin{array}{c}
\left(X_{i}\sigma^{i}+X_{0}\right)\left|\psi^{B}\right\rangle =0\\
\left(X_{i}\sigma^{i}-X_{0}\right)\left|\psi^{A}\right\rangle =0
\end{array}\right.\label{eq:46}
\end{eqnarray}
The projective Hilbert space for two non-entangled qubits A and B is expected
to be the product of two 2-dimensional spheres $S_{A}^{2}\times S_{B}^{2}$,
each sphere being the Bloch sphere associated with the given qubit \cite{Mosseri - Dandoloff}.
This property is clearly displayed here with the re-production of
the Cartan-Weyl equations (\ref{eq:19}). Equations (\ref{eq:41})
can then be written in the form 
\begin{eqnarray}
\left\{ \begin{array}{c}
\left(X_{i}^{B}\sigma^{i}+X_{0}^{B}\right)\left|\psi^{B}\right\rangle =-M\, e^{i\frac{\omega}{2}}\left|\psi^{A}\right\rangle \\
\left(X_{i}^{A}\sigma^{i}-X_{0}^{A}\right)e^{i\frac{\omega}{2}}\left|\psi^{A}\right\rangle =M\left|\psi^{B}\right\rangle 
\end{array}\right.\label{eq:47}
\end{eqnarray}
indicating that $M$ mixes the two single qubit states $\left|\psi^{A,B}\right\rangle $
into an entangled two qubit state $\left|\Psi\right\rangle $, in
a similar way to the mixing of the left- and right-chiral Weyl spinors
by mass terms into a propagating Dirac spinor.
The ``entanglement'', we invoke, arises as the term which does not allow a generalized
Bloch sphere to be given from the sum of two Bloch spheres coordinates. Below (see Sec.\ref{mixing in qubit geometry}) 
we will associate this
geometrical ``entanglement'' with the standard entanglement from quantum information theory.

\section{Decoupling of generalized Bloch sphere transformations}

Under an orthogonal transformation $L_{l}^{k}\in SO\left(5,1\right)$
of generalized Bloch sphere coordinates a two-qubits state $\left|\Psi\right\rangle $
transforms according to
\begin{equation}
\left|\Psi\right\rangle '\left(L_{l}^{k}X^{l}\right)=\Lambda\left|\Psi\right\rangle \left(X^{k}\right),\qquad k,l=0,1,2,3,4,5
\end{equation}
where 
\begin{equation}
\Lambda=\exp\left(\frac{1}{2}\omega_{kl}\mathcal{M}^{kl}\right)=\exp\left(\frac{i}{8}\omega_{kl}\left[\gamma^{k},\gamma^{l}\right]\right)
\end{equation}
the invertible elements that define $Spin\left(5,1\right)$ group,
the double cover of $SO\left(5,1\right)$. The generators $\mathcal{M}^{kl}$,
of rotations in $Spin\left(5,1\right)$, satisfy the Lie algebra of
$SO\left(5,1\right)$:
\begin{equation}
\left[\mathcal{M}^{kl},\mathcal{M}^{mn}\right]=i\left(\mathcal{M}^{kn}\mathrm{h}^{lm}+\mathcal{M}^{lm}\mathrm{h}^{kn}-\mathcal{M}^{km}\mathrm{h}^{ln}-\mathcal{M}^{ln}\mathrm{h}^{km}\right)
\end{equation}
with $\mathrm{h}=\text{diag}\left(-,+,+,+,+,+\right)$ the corresponding
metric. The algebra of these generators cannot decouple into two algebras,
one for each set of generators associated with single qubits.

Let us set 
\begin{equation}
X_{4}=X_{5}=0
\end{equation}
in order to obtain the projection: 
\begin{equation}
X_{1}^{2}+X_{2}^{2}+X_{3}^{2}-X_{0}^{2}=0
\end{equation}
of generalized Bloch sphere (\ref{eq:39}). In that case, as we have
seen, the generators $\mathcal{M}^{kl}$ reduce to 
\begin{equation}
M^{\mu\nu}=\frac{i}{4}\left[\gamma^{\mu},\gamma^{\nu}\right],\qquad\mu,\nu=0,1,2,3
\end{equation}
and satisfy the Lie algebra of $SO\left(3,1\right)$:
\begin{equation}
\left[M^{\kappa\lambda},M^{\mu\nu}\right]=i\left(M^{\kappa\nu}\mathrm{h}^{\lambda\mu}+M^{\lambda\mu}\mathrm{h}^{\kappa\nu}-M^{\kappa\mu}\mathrm{h}^{\lambda\nu}-M^{\lambda\nu}\mathrm{h}^{\kappa\mu}\right)
\end{equation}
with $\mathrm{h}=\text{diag}\left(-,+,+,+\right)$ the reduced metric.
Due to the reducible representation of generators $\gamma^{\mu}$
(see \ref{sub:Left-handed-and-right-handed}), $M^{\mu\nu}$ split
into two pieces
\begin{equation}
M_{L}^{\mu\nu}\equiv\frac{i}{4}\left[\zeta^{\mu},\theta^{\nu}\right]\text{ and }M_{R}^{\mu\nu}\equiv\frac{i}{4}\left[\theta^{\mu},\zeta^{\nu}\right].
\end{equation}
We have to show now that these two generators decouple into two $SU\left(2\right)$
algebras, one for each single qubit. Starting with generator $M_{L}^{\mu\nu}$,
one can explicitly separates it into two generators
\begin{equation}
\begin{aligned}J_{i}^{L}= & \frac{1}{2}\varepsilon_{ijk}M_{jk}^{L}=\frac{1}{2}\sigma_{i}\\
K_{i}^{L}= & M_{0i}^{L}=-\frac{i}{2}\sigma_{i}
\end{aligned}
\end{equation}
which are respectively the generator of rotations of Bloch sphere
and of Bloch sphere resizing. The latter is possible due to the not normalized quantum states we used. 
We define the combinations of these two
sets of generators 
\begin{equation}
\begin{aligned}\mathcal{L}_{i}= & \frac{1}{2}\left(J_{i}^{L}+iK_{i}^{L}\right)=\frac{1}{2}\sigma_{i}\\
\bar{\mathcal{L}}_{i}= & \frac{1}{2}\left(J_{i}^{L}-iK_{i}^{L}\right)=0
\end{aligned}
\end{equation}
and find 
\begin{equation}
\begin{aligned}\left[\mathcal{L}_{i},\mathcal{L}_{j}\right]= & i\varepsilon_{ijk}\mathcal{L}_{k}\\
\left[\mathcal{L}_{i},\mathcal{\bar{L}}_{j}\right]= & 0
\end{aligned}
\end{equation}

Similarly with generators $M_{R}^{\mu\nu}$ we obtain

\begin{equation}
\begin{aligned}\mathcal{R}_{i}= & \frac{1}{2}\left(J_{i}^{R}+iK_{i}^{R}\right)=0\\
\bar{\mathcal{R}}_{i}= & \frac{1}{2}\left(J_{i}^{R}-iK_{i}^{R}\right)=\frac{1}{2}\sigma_{i}
\end{aligned}
\end{equation}
which satisfy the commutation relations
\begin{equation}
\begin{aligned}\left[\bar{\mathcal{R}}_{i},\bar{\mathcal{R}}_{j}\right]= & i\varepsilon_{ijk}\mathcal{\bar{R}}_{k}\\
\left[\mathcal{R}_{i},\mathcal{\bar{\mathcal{R}}}_{j}\right]= & 0.
\end{aligned}
\end{equation}
Therefore the generators, associated with generalized Bloch sphere,
can be split into two subsets, which commute with each other and each
satisfy the same commutation relations as the group $SU\left(2\right)$,
provided that the number of generalized Bloch sphere dimensions is
reduced by two.

\section{Mixing in qubit geometry\label{mixing in qubit geometry}}

The Hilbert space ${\cal E}$ for the system of two qubits is the
tensor product of the individual Hilbert spaces ${\cal E}_{1}\otimes{\cal E}_{2}$,
with a direct product basis $\{\left|00\right\rangle ,\left|01\right\rangle ,\left|10\right\rangle ,\left|11\right\rangle \}$.
Thus, the two qubit state $\left|\Psi\right\rangle $ reads
\begin{equation}
\left|\Psi\right\rangle =a\left|00\right\rangle +b\left|01\right\rangle +c\left|10\right\rangle +d\left|11\right\rangle 
\end{equation}
$\left|\Psi\right\rangle $ is said ``separable'' if, at the price
of possible basis changes in ${\cal E}_{1}$ and ${\cal E}_{2}$ separately,
it can be written as a single product. As is well known, the separability
condition reads: 
\begin{equation}
ad-bc=0.\label{eq:50}
\end{equation}

For normalized $\left|\Psi\right\rangle$, the normalization condition $\left|a\right|^2+\left|b\right|^2+\left|c\right|^2+\left|d\right|^2=1$
identifies ${\cal E}$ to the $7$-dimensional sphere $S^7$. Using quaternions it can be introduced the non-trivial 
fibration $S^7 \overset{S^3}{\rightarrow} S^4$, with fibers $S^3$ and base $S^4$ \cite{Mosseri - Dandoloff}. The base $S^4$ is the projective 
Hilbert space of ${\cal E}$ and constitutes the generalized Bloch sphere \footnote{representing the state $\left|\Psi\right\rangle$
on $S^7$ as a pair of quaternions, the $S^7$ points $(q_1,q_2)$ and $(q_{1}q,q_{2}q)$, with $q$ a unit quaternion (defines
$S^3$)  are equivalence classes.}
As for the standard Bloch sphere case, 
the coordinates of $S^4$ are also expectation values of simple operators in the two-qubits state.
Two coordinates are recovered by an operator, termed ``entanglor'' \cite{Mosseri - Dandoloff}, which provides an estimate for the entanglement,
through a quantity called the ``concurrence'' \cite{Wootters}.

The aim of this paragraph is to establish a correspondance between
the well-known separability condition (\ref{eq:50}) in qubit geometry
and the non mixing criteria (\ref{eq:45}) in spinor geometry we obtained
at the previous paragraph. Let us start with the generalized Bloch
sphere: 
\begin{equation}
\mathcal{X}_{1}^{2}+\mathcal{X}_{2}^{2}+\mathcal{X}_{3}^{2}+\mathcal{X}_{4}^{2}+\mathcal{X}_{5}^{2}=\mathcal{X}_{0}^{2}\label{eq:51}
\end{equation}
which provides an, up to a global phase, representation of:
\begin{equation}
\left|\widetilde{\Psi}\right\rangle :=\left(\begin{array}{cccc}
J & 0 & 0 & 0\\
0 & 1 & 0 & 0\\
0 & 0 & J & 0\\
0 & 0 & 0 & 1
\end{array}\right)\left|\Psi\right\rangle =\left(\begin{array}{c}
a^{*}\\
b\\
c^{*}\\
d
\end{array}\right)\label{eq:59}
\end{equation}
with components, given by:
\begin{equation}
\begin{aligned}\mathcal{X}_{0}= & \left\langle \widetilde{\Psi}\right|\left.\widetilde{\Psi}\right\rangle =\left|a\right|^{2}+\left|b\right|^{2}+\left|c\right|^{2}+\left|d\right|^{2}\\
\mathcal{X}_{1}= & \left\langle \widetilde{\Psi}\right|\mathcal{G}_{1}\left|\widetilde{\Psi}\right\rangle =2\, Re\left(ab+d^{*}c^{*}\right)\\
\mathcal{X}_{2}= & \left\langle \widetilde{\Psi}\right|\mathcal{G}_{2}\left|\widetilde{\Psi}\right\rangle =2\, Im\left(ab+d^{*}c^{*}\right)\\
\mathcal{X}_{3}= & \left\langle \widetilde{\Psi}\right|\mathcal{G}_{3}\left|\widetilde{\Psi}\right\rangle =\left|a\right|^{2}-\left|b\right|^{2}+\left|c\right|^{2}-\left|d\right|^{2}\\
\mathcal{X}_{4}= & \left\langle \widetilde{\Psi}\right|\mathcal{G}_{4}\left|\widetilde{\Psi}\right\rangle =2\, Im\left(cb-ad\right)\\
\mathcal{X}_{5}= & \left\langle \widetilde{\Psi}\right|\mathcal{G}_{5}\left|\widetilde{\Psi}\right\rangle =2\, Re\left(cb-ad\right)
\end{aligned}\label{eq:63}
\end{equation}
the expectation values of operators: 
\begin{equation}
\begin{aligned}\mathcal{G}_{1}= & \sigma_{1}\otimes\mathbf{1}_{2}=\left(\begin{array}{cccc}
0 & 1 & 0 & 0\\
1 & 0 & 0 & 0\\
0 & 0 & 0 & 1\\
0 & 0 & 1 & 0
\end{array}\right) & \mathcal{G}_{2}= & \sigma_{2}\otimes\sigma_{3}=\left(\begin{array}{cccc}
0 & -i & 0 & 0\\
i & 0 & 0 & 0\\
0 & 0 & 0 & i\\
0 & 0 & -i & 0
\end{array}\right)\\
\mathcal{G}_{3}= & \sigma_{3}\otimes\mathbf{1}_{2}=\left(\begin{array}{cccc}
1 & 0 & 0 & 0\\
0 & -1 & 0 & 0\\
0 & 0 & 1 & 0\\
0 & 0 & 0 & -1
\end{array}\right) & \mathcal{G}_{4}= & \sigma_{2}\otimes\sigma_{1}=\left(\begin{array}{cccc}
0 & 0 & 0 & -i\\
0 & 0 & i & 0\\
0 & -i & 0 & 0\\
i & 0 & 0 & 0
\end{array}\right)\\
\mathcal{G}_{5}= & \sigma_{2}\otimes\sigma_{2}=\left(\begin{array}{cccc}
0 & 0 & 0 & -1\\
0 & 0 & 1 & 0\\
0 & 1 & 0 & 0\\
-1 & 0 & 0 & 0
\end{array}\right)
\end{aligned}
\end{equation}
which coincide with the generators of Clifford algebra $\mathcal{C}l\left(5\right)$ since they satisfy:
\begin{equation}
\left\{ \mathcal{G}_{t},\mathcal{G}_{p}\right\} =2\delta_{tp}.
\end{equation}

The reason that the radius of eq.(\ref{eq:51}) in not 1 is that the state $\left|\widetilde{\Psi}\right\rangle$ is not
normalized.
The choice of $\left|\widetilde{\Psi}\right\rangle$, eq.(\ref{eq:59}), also needs an explanation. The operator $J$, which takes the complex conjugate of a complex
number, is an essential ingredient in the definition of the operator called ``entanglor'' (see \cite{Mosseri - Dandoloff} for more 
details), we mention before. Here, we take the operators, which recover the generalized Bloch sphere coordinates,  
to coincide with the generators of Clifford algebra $\mathcal{C}l\left(5\right)$, thus $J$ is applied directly to $\left|\Psi\right\rangle$.
The Cartan's equation associated to quantum state $\left|\widetilde{\Psi}\right\rangle$ takes the form:
\begin{equation}
\left(\mathcal{G}^{t}\mathcal{X}_{t}-\mathcal{X}_{0}\right)\left|\widetilde{\Psi}\right\rangle =0.
\end{equation}
Note that the separability condition (\ref{eq:50}) implies 
\begin{equation}
\mathcal{X}_{4}=\mathcal{X}_{5}=0,
\end{equation}
thus, indeed, the dimensions of two qubits Bloch sphere are entanglement
sensitive. 

Let us now examine the effects of these results on the quantity 
\begin{equation}
 M=\sqrt{\mathcal{X}_{4}^2+\mathcal{X}_{5}^2}=2\left|ad-cb\right|\label{65}
\end{equation}
we met
at subsection \ref{mixing terms}. According to eqs.(\ref{eq:63}), $M$ vanishes for non entangled states, and takes
its maximal norm for maximally entangled states. In addition, due to eq.(\ref{65}) $M$ is equal to the concurrence
$C$ \cite{Wootters}, $M=C$. 

\section{The Cartan equation of four-qubits state}

In this section we apply our imbedding formalism to the case of four-qubits state. We consider four ``separable''
single qubit states $\left|\psi^A\right\rangle,\left|\psi^B\right\rangle,\left|\phi^C\right\rangle$ and $\left|\phi^D\right\rangle$,
in the sense we mention at section \ref{two-qubits}, which are associated with the Cartan's equations
\begin{equation}
\begin{aligned}\left(X_{i}^{A}\sigma^{i}-X_{0}^{A}\right)\left|\psi^{A}\right\rangle  & =0\\
\left(X_{i}^{B}\sigma^{i}+X_{0}^{B}\right)\left|\psi^{B}\right\rangle  & =0\\
\left(Y_{i}^{C}\sigma^{i}-Y_{0}^{C}\right)\left|\phi^{C}\right\rangle  & =0\\
\left(Y_{i}^{D}\sigma^{i}+Y_{0}^{D}\right)\left|\phi^{D}\right\rangle  & =0
\end{aligned}
\label{eq:7.87}
\end{equation}
respectively.

According to section \ref{two-qubits}, the coupling of two single qubits states generates a quadratic equation which 
is associated to the generalized Bloch sphere. Let us apply here this coupling formula and take two quadratic
equations, due to the two couples of single qubit states
\begin{eqnarray}
 X_1^2+X_2^2+X_3^2+X_4^2+X_5^2 -X_0^2 &=& 0,\\
 Y_1^2+Y_2^2+Y_3^2+Y_4^2+Y_5^2 -Y_0^2&=& 0,
\end{eqnarray}
where the first is associated to state $\left|\Psi\right\rangle=\left|\psi^A\right\rangle\oplus\left|\psi^B\right\rangle$ and the second to state 
$\left|\Phi\right\rangle=\left|\phi^C\right\rangle\oplus\left|\phi^D\right\rangle$. According to the proposition 1 
and the properties of Weyl spinors (see Sec. \ref{introd. pure spinor}), it may be easily shown that the components $X_{\rho},Y_{\rho}$
of null vectors $X,Y$ can be expressed in the following form:
\begin{equation}
 Z_{\rho}^{\pm}=\left\langle\bar{\Omega}\right|\Gamma_{\rho}\left(\frac{1\pm\Gamma_7}{2}\right)\left|\Omega\right\rangle\qquad \rho=0,1,2,3,4,5
\end{equation}
where $Z_{\rho}^{+}\equiv X_{\rho},Z_{\rho}^{-}\equiv Y_{\rho}$ and 
$\left|\Omega\right\rangle=\left|\Psi\right\rangle\oplus\left|\Phi\right\rangle$ with 
$\left\langle\bar{\Omega}\right|=\left|\Omega\right\rangle^{\dagger}\Gamma_0$. $\Gamma_{\rho}$, having the form:
\begin{equation}
\Gamma_{\rho}:\,\Gamma_{a}=\left(\begin{array}{cc}
0 & \gamma_{a}\\
\gamma_{a} & 0
\end{array}\right),\;\Gamma_{4}=\left(\begin{array}{cc}
0 & -i\mathbf{1}_{4}\\
i\mathbf{1}_{4} & 0
\end{array}\right),\qquad a=0,1,2,3,5
\end{equation}
satisfy the equation:
\begin{equation}
\left\{ \Gamma_{\rho},\Gamma_{\sigma}\right\} =2g_{\rho\sigma}=2\left(\begin{array}{cccccc}
-1 & 0 & 0 & 0 & 0 & 0\\
0 & 1 & 0 & 0 & 0 & 0\\
0 & 0 & 1 & 0 & 0 & 0\\
0 & 0 & 0 & 1 & 0 & 0\\
0 & 0 & 0 & 0 & 1 & 0\\
0 & 0 & 0 & 0 & 0 & 1
\end{array}\right)
\end{equation}
The volume element $\Gamma_7$ is given by:
\begin{equation}
 \Gamma_7=\Gamma_0\Gamma_1\Gamma_2\Gamma_3\Gamma_4\Gamma_5=\left(\begin{array}{cc}
                                                                  \mathbf{1}_{4} & 0\\
                                                                  0 & -\mathbf{1}_{4}
                                                                 \end{array}\right)
\end{equation}

Now let us sum the real null vectors $Z_{\rho}^{\pm}$ and take:
\begin{equation}
 Z_{\rho}=Z_{\rho}^{+}+Z_{\rho}^{-}=\left\langle\bar{\Omega}\right|\Gamma_{\rho}\left|\Omega\right\rangle
\end{equation}
Then, because of Proposition 2, together with the components 
\begin{equation}
 Z_6=i\left\langle\bar{\Omega}\right.\left|\Omega\right\rangle,\qquad Z_7=\left\langle\bar{\Omega}\right|\Gamma_{7}\left|\Omega\right\rangle
\end{equation}
they build up, for $\left|\Omega\right\rangle$ a pure spinor of $\mathcal{C}l\left(7,1\right)$, a real null vector
$Z\in \mathbb{R}^{7,1}$ with components:
\begin{equation}
 Z_{P}=\left\langle\bar{\Omega}\right|\Gamma_{P}\left|\Omega\right\rangle \quad \text{with} \quad \Gamma_{P}=\left\{\Gamma_{\rho},i,\Gamma_{7}\right\}\quad A=0,1,\ldots,7.\label{eq:7.96}
\end{equation}
It may be easily verified that, in accordance with Proposition 2, $Z_P$ given by (\ref{eq:7.96}) satisfies to 
\begin{equation}
 Z_1^2+Z_2^2+Z_3^2+Z_4^2+Z_5^2+Z_6^2+Z_7^2-Z_0^2=0.
\end{equation}

The Cartan's equation for $\left|\Omega\right\rangle$, taking into account of (\ref{eq:7.96}) is:
\begin{equation}
 \left(Z_{\rho}\Gamma^{\rho}+iZ_6+Z_7\Gamma^7\right)\left|\Omega\right\rangle=0,
\end{equation}
or in block notation
\begin{multline}
 \left(\begin{array}{cccc}
        \left(Z_7+iZ_6\right)\mathbf{1}_{2} & 0 & \left(Z_5-iZ_4\right)\mathbf{1}_{2} & \left(Z_i\sigma^i+Z_0\right)\mathbf{1}_{2}\\
        0 & \left(Z_7+iZ_6\right)\mathbf{1}_{2} & \left(Z_i\sigma^i-Z_0\right)\mathbf{1}_{2} & \left(-Z_5-iZ_4\right)\mathbf{1}_{2}\\
        \left(Z_5+iZ_4\right)\mathbf{1}_{2} & \left(Z_i\sigma^i+Z_0\right)\mathbf{1}_{2} & \left(-Z_7+iZ_6\right)\mathbf{1}_{2} & 0\\
        \left(Z_i\sigma^i-Z_0\right)\mathbf{1}_{2} & \left(-Z_5+iZ_4\right)\mathbf{1}_{2} & 0 & \left(-Z_7+iZ_6\right)\mathbf{1}_{2}
       \end{array}\right)\times\\
       \times\left(\begin{array}{c}
              \left|\psi^A\right\rangle\\
              \left|\psi^B\right\rangle\\
              \left|\phi^C\right\rangle\\
              \left|\phi^D\right\rangle\\
             \end{array}\right)
             =0,
\qquad i=1,2,3
\end{multline}
or
\begin{equation}
 \left\{
 \begin{array}{c}
  \left(Z_i\sigma^i-Z_0\right)\left|\psi^A\right\rangle=\left(Z_5-iZ_4\right)\left|\psi^B\right\rangle + \left(Z_7-iZ_6\right)\left|\phi^D\right\rangle\\
  \left(Z_i\sigma^i+Z_0\right)\left|\psi^B\right\rangle=-\left(Z_5+iZ_4\right)\left|\psi^A\right\rangle + \left(Z_7-iZ_6\right)\left|\phi^C\right\rangle\\
  \left(Z_i\sigma^i-Z_0\right)\left|\phi^C\right\rangle=\left(Z_5+iZ_4\right)\left|\phi^D\right\rangle - \left(Z_7+iZ_6\right)\left|\psi^B\right\rangle\\
  \left(Z_i\sigma^i+Z_0\right)\left|\phi^D\right\rangle=\left(Z_5-iZ_4\right)\left|\phi^C\right\rangle - \left(Z_7+iZ_6\right)\left|\psi^A\right\rangle
 \end{array}\right.\label{eq:7.100}
\end{equation}
Let us express the complex vectors $Z_5+iZ_4$ and $Z_7+iZ_6$ in the polar form:
\begin{eqnarray}
 Z_5\pm iZ_4&=&M_1\exp\left(\pm i\frac{\omega_1}{2}\right) \\ 
 Z_7\pm iZ_6&=&M_2\exp\left(\pm i\frac{\omega_2}{2}\right)
\end{eqnarray}
Substituting this in eq.(\ref{eq:7.100}) and multiplying the first and fourth by $\exp\left(i\tfrac{\omega_1}{2}\right)$
and afterwards the first and second  by $\exp\left(i\tfrac{\omega_2}{2}\right)$ we obtain:
\begin{equation}
 \left\{
 \begin{array}{rcl}
  \left(Z_i\sigma^i-Z_0\right)\exp\left(i\frac{\omega_2+\omega_2}{2}\right)\left|\psi^A\right\rangle &=&M_1\exp\left(i\frac{\omega_2}{2}\right)\left|\psi^B\right\rangle + M_2\exp\left(i\frac{\omega_1}{2}\right)\left|\phi^D\right\rangle\\
  \left(Z_i\sigma^i+Z_0\right)\exp\left(i\frac{\omega_2}{2}\right)\left|\psi^B\right\rangle&=&-M_1\exp\left(i\frac{\omega_1 +\omega_2}{2}\right)\left|\psi^A\right\rangle + M_2\left|\phi^C\right\rangle\\
  \left(Z_i\sigma^i-Z_0\right)\left|\phi^C\right\rangle&=&M_1\exp\left(i\frac{\omega_1}{2}\right)\left|\phi^D\right\rangle - M_2\exp\left(i\frac{\omega_2}{2}\right)\left|\psi^B\right\rangle\\
  \left(Z_i\sigma^i+Z_0\right)\exp\left(i\frac{\omega_1}{2}\right)\left|\phi^D\right\rangle&=&M_1\left|\phi^C\right\rangle - M_2\exp\left(i\frac{\omega_1 +\omega_2}{2}\right)\left|\psi^A\right\rangle
 \end{array}\right.\label{eq:7.103}
\end{equation}
or setting
\begin{eqnarray}
 \left|\psi^A{'}\right\rangle&:=&\exp\left(i\frac{\omega_2+\omega_2}{2}\right)\left|\psi^A\right\rangle \nonumber\\ \nonumber
 \left|\psi^B{'}\right\rangle&:=&\exp\left(i\frac{\omega_2}{2}\right)\left|\psi^B\right\rangle\\
 \left|\phi^D{'}\right\rangle&:=&\exp\left(i\frac{\omega_1}{2}\right)\left|\phi^D\right\rangle \nonumber
\end{eqnarray}
eqs.(\ref{eq:7.103}) become
\begin{equation}
 \left\{
 \begin{array}{rcl}
  \left(Z_i\sigma^i-Z_0\right)\left|\psi^A{'}\right\rangle &=&M_1\left|\psi^B{'}\right\rangle + M_2\left|\phi^D{'}\right\rangle\\
  \left(Z_i\sigma^i+Z_0\right)\left|\psi^B{'}\right\rangle&=&-M_1\left|\psi^A{'}\right\rangle + M_2\left|\phi^C\right\rangle\\
  \left(Z_i\sigma^i-Z_0\right)\left|\phi^C\right\rangle&=&M_1\left|\phi^D{'}\right\rangle - M_2\left|\psi^B{'}\right\rangle\\
  \left(Z_i\sigma^i+Z_0\right)\left|\phi^D{'}\right\rangle&=&M_1\left|\phi^C\right\rangle - M_2\left|\psi^A{'}\right\rangle
 \end{array}\right.\label{eq:7.104}
\end{equation}
Note that setting 
\begin{equation}
 M_1=M_2=0
\end{equation}
eqs.(\ref{eq:7.104}) reduce to eqs.(\ref{eq:7.87}) indicating that $M_1,M_2$ mix the four single qubit states $\left|\psi^A\right\rangle,\left|\psi^B\right\rangle,\left|\phi^C\right\rangle,\left|\phi^D\right\rangle$
into an entangled four-qubits state $\left|\Omega\right\rangle$. As in the case of two-qubits, the quantum 
entanglement we consider in this section is defined as the way that four disjoint
Cartan-Weyl equations of single qubits are correlated. Thus, the results of this section generalize the findings of 
two-qubits case. 

Let us now associate our findings with what we already know regarding four-qubits systems. $\left|\Omega\right\rangle$,
as an 8-component spinor, cannot be associated with a general 4-qubit pure state of the form $\left|\Phi\right\rangle=a_{0}\left|0000\right\rangle+\ldots+a_{15}\left|1111\right\rangle$.
However, the formation procedure we followed to build $\left|\Omega\right\rangle$ allow us to set
\begin{eqnarray}
 \left|\Omega\right\rangle&=&a_1\left|0000\right\rangle+a_2\left|0011\right\rangle+a_3\left|0101\right\rangle+a_4\left|0110\right\rangle \\ \nonumber
                           & &+a_5\left|1001\right\rangle+a_6\left|1010\right\rangle+a_7\left|1100\right\rangle+a_8\left|1111\right\rangle
\end{eqnarray}
Explicitly, as a first level action, we ``mix'' the four
single-qubit states per pairs and take a pair of 2-qubit states. Then, as a second level action we partially ``mix'' the
pair of 2-qubit states into a 4-qubit state. Just to become the derivation of this 4-qubit state more clear see (\ref{eq:7.107}):
\begin{equation}
   \left.\begin{array}{c}
           \left.\begin{array}{c}
            \left\{\left|0\right\rangle,\left|1\right\rangle\right\} \\
            \left\{\left|0\right\rangle,\left|1\right\rangle\right\}
           \end{array}\right\}\left\{\{\left|00\right\rangle,\left|11\right\rangle\},\{\left|01\right\rangle,\left|10\right\rangle\}\right\} \\
           
           \left.\begin{array}{c}
            \left\{\left|0\right\rangle,\left|1\right\rangle\right\} \\
            \left\{\left|0\right\rangle,\left|1\right\rangle\right\}
           \end{array}\right\}\left\{\{\left|00\right\rangle,\left|11\right\rangle\},\{\left|01\right\rangle,\left|10\right\rangle\}\right\}
   \end{array}\right\}\begin{array}{r}
                       \left\{\left|0000\right\rangle,\left|0011\right\rangle,\left|1100\right\rangle,\left|1111\right\rangle\right. \\ 
                            \left.\left|0101\right\rangle,\left|0110\right\rangle,\left|1001\right\rangle,\left|1010\right\rangle \right\}
                      \end{array}\label{eq:7.107}
\end{equation}

Recently it was obtained a classification of 4-qubit pure states up to SLOCC equivalence \cite{Verstraete et al}. In this
work it was identified nine different families of pure states of 4 qubits generated by SLOCC operations, with only one 
family to be generic, the $G_{abcd}$ state. As we will see below, $G_{abcd}$ is connected to $\left|\Omega\right\rangle$. 

Given the state $\left|\widetilde{\Omega}\right\rangle=f(J)\left|\Omega\right\rangle$, where $f(J)$ a function of operator $J$ which takes the complex conjugate of a complex number,
 we obtain the quadratic equation
\begin{equation}
 \mathcal{Z}_1^2+\mathcal{Z}_2^2+\mathcal{Z}_3^2+\mathcal{Z}_4^2+\mathcal{Z}_5^2+\mathcal{Z}_6^2+\mathcal{Z}_7^2=\mathcal{Z}_0^2
\end{equation}
with components, given by:
\begin{equation}
\begin{aligned}\mathcal{Z}_{0}= & \left\langle\widetilde{\Omega}\right|\left.\widetilde{\Omega}\right\rangle =\left|a_1\right|^{2}+\left|a_2\right|^{2}+\left|a_3\right|^{2}+\left|a_4\right|^{2}+\left|a_5\right|^{2}+\left|a_6\right|^{2}+\left|a_7\right|^{2}+\left|a_8\right|^{2}\\
\mathcal{Z}_{1}= & \left\langle\widetilde{\Omega}\right|\mathcal{J}_{1}\left|\widetilde{\Omega}\right\rangle =2\, Re\left(a_2a_1+a_3^{*}a_4^{*}+a_6^{*}a_5^{*}+a_7a_8\right)\\
\mathcal{Z}_{2}= & \left\langle\widetilde{\Omega}\right|\mathcal{J}_{2}\left|\widetilde{\Omega}\right\rangle =2\, Im\left(a_2a_1+a_3^{*}a_4^{*}+a_6^{*}a_5^{*}+a_7a_8\right)\\
\mathcal{Z}_{3}= & \left\langle\widetilde{\Omega}\right|\mathcal{J}_{3}\left|\widetilde{\Omega}\right\rangle =\left|a_1\right|^{2}-\left|a_2\right|^{2}+\left|a_3\right|^{2}-\left|a_4\right|^{2}+\left|a_5\right|^{2}-\left|a_6\right|^{2}+\left|a_7\right|^{2}-\left|a_8\right|^{2} \\
\mathcal{Z}_{4}= & \left\langle\widetilde{\Omega}\right|\mathcal{J}_{4}\left|\widetilde{\Omega}\right\rangle =2\, Re\left(-a_1a_4+a_2a_3+a_8a_5-a_7a_6\right)\\
\mathcal{Z}_{5}= & \left\langle\widetilde{\Omega}\right|\mathcal{J}_{5}\left|\widetilde{\Omega}\right\rangle =2\, Im\left(-a_1a_4+a_2a_3+a_8a_5-a_7a_6\right)\\
\mathcal{Z}_{6}= & \left\langle\widetilde{\Omega}\right|\mathcal{J}_{6}\left|\widetilde{\Omega}\right\rangle =2\, Re\left(-a_8a_1+a_7a_2+a_6a_3-a_5a_4\right)\\
\mathcal{Z}_{7}= & \left\langle\widetilde{\Omega}\right|\mathcal{J}_{7}\left|\widetilde{\Omega}\right\rangle =2\, Im\left(-a_8a_1+a_7a_2+a_6a_3-a_5a_4\right)
\end{aligned}\label{eq:109}
\end{equation}
the expectation values of operators:
\begin{eqnarray}
\mathcal{J}_{1}&=& \sigma_{1}\otimes\mathbf{1}_{2}\otimes\mathbf{1}_{2} \nonumber \\
\mathcal{J}_{2}&=& \sigma_{2}\otimes\sigma_{3}\otimes\mathbf{1}_{2}\\
\mathcal{J}_{3}&=& i\mathcal{J}_{1}\mathcal{J}_{2}\mathcal{J}_{3}\mathcal{J}_{4}\mathcal{J}_{5}\mathcal{J}_{6} \nonumber \\
\mathcal{J}_{4}&=& \sigma_{2}\otimes\sigma_{2}\otimes\sigma_{3} \nonumber \\
\mathcal{J}_{5}&=& \sigma_{2}\otimes\sigma_{1}\otimes\mathbf{1}_{2} \nonumber \\
\mathcal{J}_{6}&=& \sigma_{2}\otimes\sigma_{2}\otimes\sigma_{1} \nonumber \\
\mathcal{J}_{7}&=& \sigma_{2}\otimes\sigma_{2}\otimes\sigma_{2}\nonumber 
\end{eqnarray}
which coincide with the generators of Clifford algebra $\mathcal{C}l\left(7\right)$ since they satisfy:
\begin{equation}
\left\{ \mathcal{J}_{k},\mathcal{J}_{l}\right\} =2\delta_{kl}\qquad k,l=1,2,\ldots,7.
\end{equation}
Thus, the state $\left|\widetilde{\Omega}\right\rangle$ is associated to Cartan's equation:
\begin{equation}
 \left(\mathcal{Z}_{k}\mathcal{J}^{k}-\mathcal{Z}_{0}\right)\left|\widetilde{\Omega}\right\rangle=0.
\end{equation}

Let us now examine the quantities $M_1$ and $M_2$ and the effects they have on state $\left|\widetilde{\Omega}\right\rangle$.
Both quantities can be written in terms 
of $\left|\widetilde{\Omega}\right\rangle$'s components, $a_1,\ldots,a_8$:
\begin{equation}
 \begin{aligned}
  M_1=\sqrt{\mathcal{Z}_{4}^2+\mathcal{Z}_{5}^2}=& 2\left|a_1a_4-a_2a_3-a_8a_5+a_7a_6\right| \\
  M_2=\sqrt{\mathcal{Z}_{6}^2+\mathcal{Z}_{7}^2}=& 2\left|a_8a_1-a_7a_2-a_6a_3+a_5a_4\right|
 \end{aligned}
\end{equation}
Starting with $M_2$, and setting
\begin{equation}
 \begin{aligned}
  a_8a_1=&\left(\frac{a+d}{2}\right)^2\\
  -a_7a_2=&\left(\frac{a-d}{2}\right)^2\\
  -a_6a_3=&\left(\frac{b+c}{2}\right)^2\\
  a_5a_4=&\left(\frac{b-c}{2}\right)^2\\
 \end{aligned}\label{eq:7-114}
\end{equation}
where $a,b,c,d$ complex numbers, we are finding that $M_2$ is equal to the entanglement monotone 
$\mathcal{M}_2\left(\left|\widetilde{\Omega}\right\rangle\right)$ \cite{Verstraete et al}, $M_2=\mathcal{M}_2\left(\left|\widetilde{\Omega}\right\rangle\right)$.
Taking into account the transformations (\ref{eq:7-114}) the state $\left|\Omega\right\rangle$ becomes
\begin{eqnarray}
 \left|\Omega\right\rangle&=&\frac{a+d}{2}\left(\left|0000\right\rangle+\left|1111\right\rangle\right)+\frac{a-d}{2}\left(\left|0011\right\rangle+\left|1100\right\rangle\right) \\ \nonumber
                           & &\frac{b+c}{2}\left(\left|0101\right\rangle+\left|1010\right\rangle\right)+\frac{b-c}{2}\left(\left|0110\right\rangle+\left|1001\right\rangle\right)
\end{eqnarray}
which, according to the classification of Verstraete et al. \cite{Verstraete et al}, defines the family of states 
$G_{abcd}$. In the light of the results of Gisin \cite{N. Gisin} and Nielsen about majorization \cite{Nielsen}, it is claimed that this
is the state with maximal 4-partite entanglement on the complete orbit generated by SLOCC operations.
$M_2$
vanishes for non entangled
states, and takes its maximal norm for maximally entangled states.

Let us next move on the quantity $M_1$. It is easily checked that, taking into account the form of $\left|\Omega\right\rangle$
and the quantity $M$, eq.(\ref{65}) from the case of two qubits, holds
\begin{equation}
   \begin{aligned}
     M_1= 2&\left|a_1a_4-a_2a_3-a_8a_5+a_7a_6\right|\leq \\
          &2\left|a_1a_4-a_2a_3\right|+2\left|-1\right|\left|a_8a_5-a_7a_6\right|=M^{(1)}+M^{(1)}
   \end{aligned}\label{eq:7-116}
 \end{equation}
 or
\begin{equation}
   \begin{aligned}
     M_1= 2&\left|a_1a_4-a_2a_3-a_8a_5+a_7a_6\right|\leq \\
          &2\left|a_1a_4-a_2a_3\right|+2\left|-1\right|\left|a_8a_5-a_7a_6\right|=M^{(2)}+M^{(2)}
   \end{aligned}\label{eq:7-117}
 \end{equation}
depending on the permutation of two EPR-pairs of qubits\footnote{The qubits 2 and 3 compose the first EPR-pair and the 
qubits 1 and 4 compose the second EPR-pair. Note the permutation invariance inside each one of EPR-pairs.}.
The symbol $M^{(i)}$ with $i=1,2$ indicates the entanglement measure (equal to the concurrence $C$) of each one of the
two complementary pairs of qubits, 
as it has been derived at section \ref{mixing terms} (see eq.(\ref{65})). 
Adding the equations (\ref{eq:7-116}) and (\ref{eq:7-117}) and dividing both sides of the new equation by 2 we take
\begin{equation}
 M_1\leq M^{(1)}+M^{(2)}
\end{equation}
Thus, we can say that the amount of entanglement of the two complementary qubit-pairs taken individually,
bounds the entanglement measured by quantity $M_1$. 
$M_1$ vanishes for states consisting of two EPR-pairs which are not entangled, 
and takes its maximal norm when these states are maximally entangled.

\section{Discussion and conclusions}

In this paper we attempted to show how the elegant geometry of pure
spinors could be helpful for throwing some light on tantalizing notion
of quantum entanglement. Our main goal was, taking advantage of the
mathematical analogy of the pure states of a two qubits system with
four-component Dirac spinors, to provide an alternative consideration
of quantum entanglement using the mathematical formulation of Cartan's
pure spinors. The understanding of the concept of entanglement applied
here is in various respect a generalized one and differs from the
one conventionally applied in the analysis of two spatially separated
spins.

A single qubit state may be represented by a two-component Pauli spinor
and, from this, if pure, vectors of null quadrics in pseudo euclidean
spaces or expectation values of the Pauli operators in our case, may be constructed. It is then
natural to try to imbed spinor spaces, and the corresponding constructed
null quadrics, in higher dimensional ones in order to derive quantum
entanglement. Starting from two pure spinor spaces we operated the
most obvious operation: their direct sum spanned by spinors of double
dimension, gave rise to a higher dimensional pure spinor space. The
corresponding generalized null vectors, which define the generalized
Bloch sphere, is given by the sum of the corresponding expectation values
and are obtained by adding two extra components. These two components
provide an entanglement measure, in the sense that setting these two
components equal to zero the generalized Bloch sphere reduces in two
Bloch spheres representing the two single qubit states.

The present explorative study based on the formal mathematical analogy
of four-component spinors with the pure states of a two qubits system
leaves many questions open for further research. It has been already
suggested that spinors provide us with a link between space-time and
entanglement properties of spinors \cite{Nicolaidis-Kiosses}. We
leave for a future publication a systematic treatment of multipartite
entanglement using pure spinors geometry. This way entanglement properties
of spinors have possibly a relation to the problem of emerging large
number of space dimensions, a subject worth of further study.

\section*{Acknowledgment}

I would like to acknowledge valuable discussions with Argyris Nicolaidis.

\end{document}